# Formation of Ideal Rashba States on Layered Semiconductor Surfaces Steered by Strain Engineering

*Wenmei Ming,[†,‡] Z. F. Wang,[†,§] Miao Zhou,[†,∥] Mina Yoon,[*‡] and Feng Liu[*†,±]*

†Department of Materials Science and Engineering, University of Utah, Salt Lake City, Utah 84112, United States

‡Center for Nanophase Materials Sciences, Oak Ridge National Laboratory, Oak Ridge, Tennessee 37831, United States

§Hefei National Laboratory for Physical Sciences at the Microscale, University of Science and Technology of China, Anhui 230026, China

∥ Key Laboratory of Optoelectronic Technology and Systems of the Education Ministry of China, College of Optoelectronic Engineering, Chongqing University, Chongqing 400044, China

±Collaborative Innovation Center of Quantum Matter, Beijing 100084, China

**ABSTACT:** Spin splitting of Rashba states in two-dimensional electron system provides a promising mechanism of spin manipulation for spintronics applications. However, Rashba states realized experimentally to date are often outnumbered by spin-degenerated substrate states at the same energy range, hindering their practical applications. Here, by density functional theory calculation, we show that Au one monolayer film deposition on a layered semiconductor surface β-InSe(0001) can possess "ideal" Rashba states with large spin splitting, which are completely situated inside the large band gap of the substrate. The position of the Rashba bands can be tuned over a wide range with respect to the substrate band edges by experimentally accessible strain. Furthermore, our nonequilibrium Green's function transport calculation shows that this system may give rise to the long-sought strong current modulation when made into a device of Datta-Das transistor. Similar systems may be identified with other metal ultrathin films and layered semiconductor substrates to realize ideal Rashba states.

Rashba states refer[1, 2] to spin splitting of two-dimensional (2D) electronic states as a result of perpendicular potential asymmetry in the presence of spin-orbit coupling (SOC). Much recent effort has been made in searching for this type of electronic states in solid state systems for potential spintronics applications, such as spin field transistor[3] and intrinsic spin Hall effect[4]. The degree of spin splitting, quantified by the Rashba parameter $\alpha$, scales with the gradient of perpendicular potential asymmetry and strength of SOC. The earliest realization of the Rashba states was made in the asymmetric quantum well formed in InGaAs/InAlAs heterostructure[5], with small spin splitting. Noble metal (e.g., Au[6], Ag[7], and Ir[8]) and *sp*-orbit heavy-metal surfaces (e.g., Bi[9], Sb[10] and Pb[11]) were shown to have large spin splitting. Heavy metal adatoms alloying with metal (e.g., Bi/Ag(111)[12]) and/or semiconductor surfaces (e.g., Bi/Si(111)[13]) were found to possess giant spin splitting. Most recently, new surface systems with Rashba states have been reported in graphene/Ni(111)[14] and molecule-adsorbed topological insulators[15-17]. However, because the substrates previously adopted are either metal or semiconductor with substantial surfaces states, typically Rashba states are overwhelmed by the large number of spin-degenerated substrate states at the same energy window, so that the desired transport of Rashba-state carriers cannot be isolated from the substrate carrier contribution, hindering their practical applications.

The effort of overcoming this limitation therefore needs to be directed to tuning the relative energy positioning between the Rashba bands and the substrate bands, in such a way that Rashba bands have significant spin splitting from the strong interaction with the substrate, while they stay separated from the substrate bands in energy. The ideal case is that the Rashba bands are completely situated inside the band gap of the substrate, so that the pure transport contributed by only Rashba states can be obtained. To the best of our knowledge, such ideal Rashba states were only realized on Te-terminated surface of BiTeX (X = I, Br and Cl)[18, 19]. In this Letter, using density functional theory (DFT) calculations, we demonstrate a new approach to realize ideal Rashba states in a hybrid system [Au/InSe(0001)] of Au one monolayer film deposition on a completely different type of substrate, that is, layered large band-gap semiconductor β-InSe(0001), which is a naturally saturated surface without any dangling bonds as encountered in conventional semiconductor surfaces. We show that the separation between Rashba bands and substrate bands can be further tuned via strain exerted on the substrate. We show potential application of our system for spin-field transistor by quantum transport calculation.

Our DFT calculations are carried out with projector augment wave pseudopotential[20] in the VASP package[21]. Perdew-Burke-Ernzerhof (PBE)[22] exchange-correlation functional and DFT-D2[23] correction of van der Waals (vdW) interaction are used for obtaining accurate geometry of layered InSe(0001). A 450 eV kinetic energy cutoff and 11×11×1 Γ centered k-mesh sampling are adopted for total energy convergence. The substrate is simulated by 7 layers slab of (1×1)-InSe(0001) with an experimental in-plane lattice constant $a$ of 4.05 Å and an vacuum layer of more than 20 Å. All the atoms are relaxed until the forces are smaller than 0.01 eV/Å. The screened HSE (Heyd, Scuseria, and Ernzerhof)[24] hybrid functional is subsequently adopted for reliable substrate band gap and positioning of Rashba bands with SOC.

Bulk β-InSe has a hexagonal layered crystal structure, with each layer consisting of Se-In-In-Se atomic planes and stacking along $z$ direction. Within each layer, In and Se are bonded covalently; between layers, they are bonded with vdW interaction. Our HSE calculation shows that InSe(0001) has a large band gap of 1.28 eV (see Figure S1 in Supporting Information), which is very close to the bulk experimental value of 1.35 eV[25]. The apparent absence of surface bands in the band gap due to the saturated layered structure of InSe(0001) is also verified from the calculated band structure. Besides, we notice that Au ultrathin films[26-28] including one monolayer thickness on InSe(0001) were successfully grown in experiment and intensively studied in photovoltaics due to the formation of abrupt Schottky interface between Au overlayer and InSe(0001), which makes our study of Au/InSe(0001) in the context of Rashba states even more relevant to experimental feasibility.

The atomic structure of Au/InSe(0001) is shown in Fig. 1(a-b) for side and top views, respectively, which consists of one monolayer Au thin film on InSe(0001). The site directly above the hollow position of both the topmost Se and In atomic planes is energetically favorable for Au adsorption. The interlayer binding energy between Au and InSe(0001) is 0.58 eV/surface-unit-cell. It indicates that the surface interaction is in an "intermediate" range[29], slightly stronger than vdW bond but significantly weaker than typical chemical bond. Figure 2(a) shows the corresponding band structures of Au/InSe(0001) along M-Γ-K directions. Bands located in the shaded areas represent the ones from substrate states, while the ones located in the gap between them are characterized as the ones from surface states. The characterization of those bands will be further discussed later. The two surface bands show a large Rashba spin splitting at the *k*-points from midpoint of M-Γ to midpoint of Γ-K and are completely inside the substrate band gap. Therefore, this system has the ideal Rashba states immune to the substrate states.

Meanwhile, in the vicinity of Γ the Rashba surface bands are only ~ 0.25 eV above the valence band maximum (VBM) of the substrate. Next, we apply both compressive and tensile biaxial strains to the substrate in order to tune the relative position of the surface bands and the substrate bands. Compressive biaxial strain is found to separate even further the surface and substrate bands, and the surface bands becomes ~ 0.40 eV above the VBM under a compressive strain of -1.2%, as shown in Fig. 2(b). At -2.4% compressive strain, the surface bands becomes ~ 0.5 eV (see Fig. S2) above VBM. It can be attributed to the substrate band gap enlargement and the down-shifting and up-shifting of VBM and conduction band maximum (CBM), respectively, relative to the surface bands (see Fig. S2). In contrast, tensile strain exerts an opposite effect that the surface and substrate bands become increasingly closer and then overlapped, as seen under a tensile strain of +2.4% in Fig. 2(c), because the band gap shrinks and the VBM up shifts and CBM down shifts.

Previous DFT and tight-binding calculations[25, 30] reported that the VBM of InSe is composed of the bonding state by $p_z$ orbitals of Se and In, and the CBM is composed of antibonding state by s and $p_z$ orbital of Se and s orbital of In. Moreover, with increasing compressive strain the energy levels of the VBM bonding state and CBM antibonding state will become lower and higher, respectively, due to decreased In-Se distance and thereby increased overlap integral between these orbitals compared with the strain-free case. This trend will be reversed under increasing tensile strain, which causes increased In-Se distance and thereby decreased overlap integral. Therefore, it is observed from our band structure results that InSe(0001) band gap increases under compressive strain and decreases under tensile strain. Besides, the interlayer Au-Se distance keeps increasing under compressive strain and decreasing under tensile strain (Fig. S3), Au onsite energy hence should be raised and lowered under compressive and tensile strains, respectively. Consequently, the Rashba bands are lifted higher under compressive strain and lower under tensile strain. Combining the three trends for VBM, CBM and Rashba bands together, it becomes clear that we can further isolate the Rashba bands from InSe(0001) with compressive strain but the opposite with tensile strain.

Because of the larger separation of the Rashba bands from VBM on -1.2% strained substrate than on strain-free case, we choose the -1.2% strained substrate with lattice constant $a$ of 4.00 Å in all the following results. We have identified the asymmetric potential responsible for the Rashba spin splitting and the nonzero Rashba parameter $\alpha \propto \int \partial V(z)/\partial z dz$[31]. The in-plane averaged potential along the stacking direction $z$: $V(z) = \frac{1}{A}\int V(x,y,z)dxdy$, where A is the surface area, is shown in Fig. 3(a). The red dashed box highlights the region of Au and the first layer of the substrate, where electrons experience an asymmetric potential. In the remaining region, however, the potential is symmetric and almost identical at each layer. This implies that the local asymmetric potential induces localized Rashba states only in the top Au + InSe layer. To confirm this, in Fig. 3(b-e) we show the local charge distribution for the Rashba states 1-4 as labeled in Fig. 2(b), at each atomic plane from the topmost Au to the bottommost Se plane. The red dashed boxes show that the charge of each state is mostly distributed in the top five atomic planes, beyond which it is negligibly small.

Microscopically, tight-binding Hamiltonian was adopted to understand Rashba spin splitting in simple systems such as graphene[32, 33] and Au(111)[34, 35], where the effect of asymmetric potential is included by adding nonzero inter-site and intra-site hopping matrix elements between orbitals, which otherwise have zero hopping matrix elements. In order to have nonzero intra-site hopping matrix elements under asymmetric potential along the $z$ direction, the Rashba wavefunction must contain orbitals with opposite parities such as s/$p_z$ and $p_z$/$d_{z2}$ at heavy atoms (See Fig. S5 and Table S1 for the analysis of intra-site s/$p_z$ hybridization leading to Rashba splitting). This is confirmed in Fig. 3(f-i) by plotting orbital compositions of the Rashba states 1-4 at Au site, which all contain large amount of $s$, and small amount of $p_z$, $p_{xy}$ and $d_{z2}$ components at Au atomic plane. While it is difficult to have a simple picture for Au/InSe(0001) on the induced intersite hopping, which in graphene and Au(111) is from nearest-neighbor σ-π hybridization between $p_{x,y}$ and $p_z$ orbitals under asymmetric potential, the existence of $p_{x,y}$ and $p_z$ orbitals at each atomic plane particularly at Se plane may still indicate the contribution from inter-site hopping to the Rashba spin splitting observed in Au/InSe(0001).

Furthermore, we show the spin polarization texture of the Rashba states in Fig. 4. The spin polarization is defined: $\vec{p}(\vec{k}) = [<S_x(\vec{k})>, <S_y(\vec{k})>, <S_z(\vec{k})>]$, where $<S_\alpha(\vec{k})> = <$

$\varphi(\vec{k})|\sigma_\alpha|\varphi(\vec{k})>$, ($\alpha = x, y, z$). It is plotted at two iso-energy (E) arcs of E = -0.28 eV and -0.40 eV. Blue and red colors represent outer and inner branches of the Rashba bands, respectively. Overall, the iso-arc is isotropic for the inner circle with smaller *k*-vector, but gains some anisotropy of hexagonal shape for the outer circle with larger *k*-vector. The *z*-component of the spin polarization is about ten times smaller than the *x*- and *y*- components, and the *xy*-plane spin polarization is nearly perpendicular to the *k*-vector. This spin polarization texture is very similar to that in 2D Rashba SOC free-electron gas, which has only in-plane spin components[36]. Because the spin polarization texture represents the effective distribution of k-dependent magnetic field generated by SOC[37], under perfect 2D Rashba SOC as in 2D Rashba SOC free-electron gas, the spin of an initially spin-up polarized electron shows sinusoidal oscillation over its travel length[38]. We thus expect that Au/InSe(0001) should possess similar spin oscillation effect. We fit our Rashba bands around Γ point to the free-electron Rashba band dispersion. The fitted Rashba parameter α is 0.45 eV Å. It corresponds to a characteristic spin precession length[39] $L_{so} = \frac{\pi\hbar^2}{2m\alpha}$ of 50 Å, over which electron up-spin is rotated to down-spin, where m is the electron effective mass. The periodicity of the sinusoidal oscillation then is $2L_{so} = 100$ Å.

Next, we proceed to show the electron spin oscillation effect in Au/InSe(0001) by calculating its spin-dependent electron conductance. Because the Rashba bands are isolated far away from the substrate band edges, we can write an effective tight-binding Hamiltonian for only the two Rashba bands, neglecting the other bands. The effective tight-binding Hamiltonian is obtained by using the two maximally localized Wannier functions as the basis set constructed with the Wannier90 package[40]. A two-terminal nanoribbon Datta-Das transistor with the width of 4*a* is shown in Fig. 5(a). The left lead is spin-up ferromagnetic and the right lead is aligned with the left lead in either parallel or antiparallel configuration[41]. We assume no SOC in the lead regions by artificially setting the spin-flip elements in the Hamiltonian matrix to zero, and the channel region has the SOC as described by the full effective Hamiltonian. Fig. 5(b) shows the band structure of the nanoribbon representing the channel region.

The spin-dependent electron ballistic conductance $G(E_F)$ of the device is calculated with nonequilibrium Green'ss function (NEGF) method in Landauer-Büttiker formalism[41]:

$$G_{pq}(E_F) = \frac{e^2}{h} Tr(\Gamma_{lp} G^R \Gamma_{rp} G^A) \qquad (1)$$

where $\Gamma$ is the interface coupling matrix between the lead and the channel and is related to the channel self-energy ($\Sigma^R$) by $\Gamma = i(\Sigma^R - \Sigma^A)$, $G^R = G^{A+}$ is retarded Green function of the channel, p, q are spin indexes and l, r denote the left and right leads, respectively. Because in this work we focus on the intrinsic modulation of the spin-polarized current by the Rashba SOC, we do not include the interfacial effects between lead and channel, disorder and/or impurities. We purposely choose the Fermi energy ($E_F$) to be at either of the two red dashed lines in Fig. 5(b), crossing only the two Rashba bands ($R_1$) when $E_F = E_1$ or two sets of Rashba bands (R1, R2) plus the spin almost degenerated bands when $E_F = E_2$.

In Fig. 5(c), $G_{\uparrow\uparrow}(E_F = E_1)$ and $G_{\uparrow\downarrow}(E_F = E_1)$ are plotted for the spin-up and spin-down conductances at the right lead, respectively, as a function of channel length (L). They both exhibit nearly perfect sinusoidal oscillation with periodicity of ~ 90$a$ (~ 360 Å). This oscillation periodicity from the nanoribbon configuration is ~ 3.6 times as large as that estimated from the 2D configuration. We confirm this increased oscillation periodicity by again fitting the nanoribbon band structure around $\Gamma$ to the free-electron Rashba band dispersion. We find that the Rashba parameter α is reduced to 0.25 eVÅ and the effective mass $m$ is reduced to be about half of that for 2D configuration. Therefore the oscillation periodicity $2L_{so}$ is increased by 3.6 times, which is consistent with the increase from the conductance calculation. We notice that previously when estimating the Rashba parameter α in 2D structure, we only include the Rashba split bands around Γ, but neglect the almost spin degenerated bands at the same energy range around M. However, when in 1D nanoribbon structure, the bands around M in 2D structure could be folded to around Γ and hence result in the modified Rashba band dispersion and the decrease of α and m. In addition, we purposely shift $E_F$ to $E_2$ and show $G_{\uparrow\uparrow}(E_F = E_2)$ and $G_{\uparrow\downarrow}(E_F = E_2)$ in Fig. 5(d). They exhibit strong oscillation but with smaller periodicity of ~ 50$a$ (~ 200 Å) compared to that when $E_F = E_1$. It is different that the oscillation does not follow well a sinusoidal curve. Noticing that Rashba bands $R_2$ have a larger Rashba parameter α of 0.45 eV Å than $R_1$, and the interband interference between $R_1$ and $R_2$ should also appear when $E_F$ crosses both of them, the overall oscillation therefore becomes less regular and has smaller periodicity.

Lastly, we point out that the compressive strain on the substrate, which is critical for improving the ideality of the Rashba bands in Au/InSe(0001), can be achieved by epitaxially growing InSe(0001) film on an additional semiconductor substrate with smaller lattice constant. Several substrates such as GaAs(111), AlAs(111), ZnSe(111) and Ge(111) all have the surface lattice constant of ~ 4.00 Å and hence impose desirable amount of compressive strain of -1.2% on InSe(0001). On the other hand, InSe(0001) can be replaced by other layered semiconductor substrates such as those from metal chalcogenides family[42] (e.g., γ-InSe, α-$In_2Se_3$, β-$In_2Se_3$, 2H-$MoS_2$), and they may also exhibit ideal Rashba states upon heavy metal deposition such as Bi, Pb, Tl, Sb and Au as the overlayers.

In conclusion, using DFT electronic structure calculations, we have demonstrated the formation of the long-sought ideal Rashba states in Au monolayer film deposition on a new type of substrate, that is, layered semiconductor surface β-InSe(0001). They are situated completely inside the substrate band gap, in the same fashion as the recently studied surface topological states[43,44], and possess large spin splitting. The ideality of the Rashba bands can be tuned over a wide range with respect to the substrate band edges with experimentally accessible strain. Furthermore, the desired property of strong modulation of spin-polarized current in Datta-Das transistor setup is confirmed by NEGF transport calculations.

**Supporting Information**

Substrate band gap, energy separation between Rashba surface bands and substrate band edges, distance between Au and InSe(0001), Rashba parameter α, band widths, Wannier fitting. This material is available free of charge via the Internet at http://pubs.acs.org.

**Corresponding Authors**

*E-mail: myoon@ornl.gov (M. Y.)

*E-mail: fliu@eng.utah.edu (F. L.)

**Notes**

The authors declare no competing financial interest.

**Acknowledgements**

The work at Utah was supported by NSF MRSEC (Grant No. DMR-1121252) (W. M. and Z. F.) and DOE-BES (Grant No. DE-FG02-04ER46148) (M. Z. and F. L.); the work at Oak Ridge National Laboratory was supported by the Laboratory Directed Research and Development

Program (W. M.) and by the Center for Nanophase Materials Sciences (M. Y.) of the Scientific User Facilities Division, Office of Basic Energy Sciences, U.S. Department of Energy. We thank the CHPC at the University of Utah for providing the computing resources. This research used resources of the National Energy Research Scientific Computing Center, supported by the Office of Science of the U.S. Department of Energy under Contract No. DE-AC02-05CH11231.

**Table of Contents Graphic**

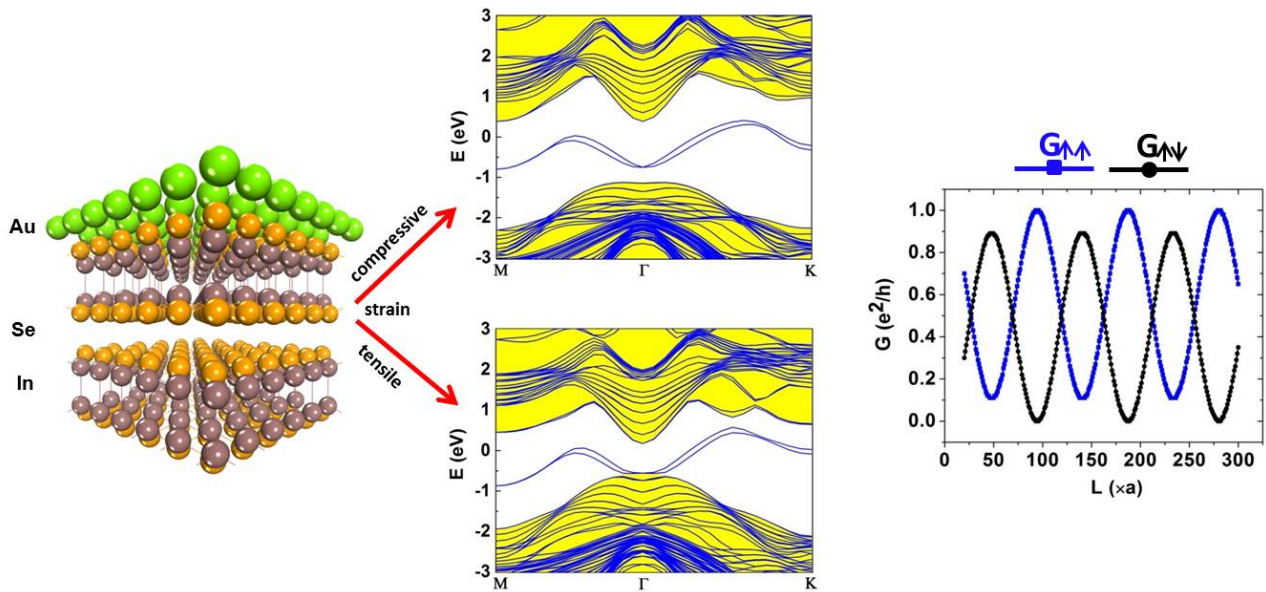

**Figures and Captions**

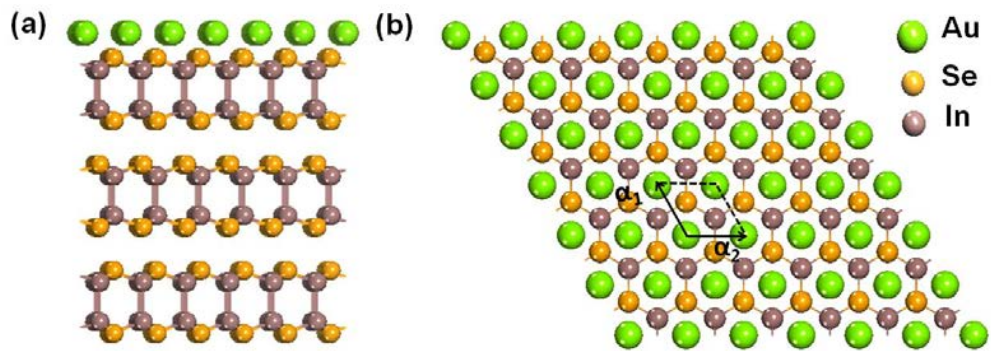

**Figure 1.** (a) Side view of the structure of Au monolayer film deposition on β-InSe(0001) substrate, which is shown with only the top three layers for clarity. (b) Top view of the structure, **α₁** and **α₂** represent the surface unit cell vectors.

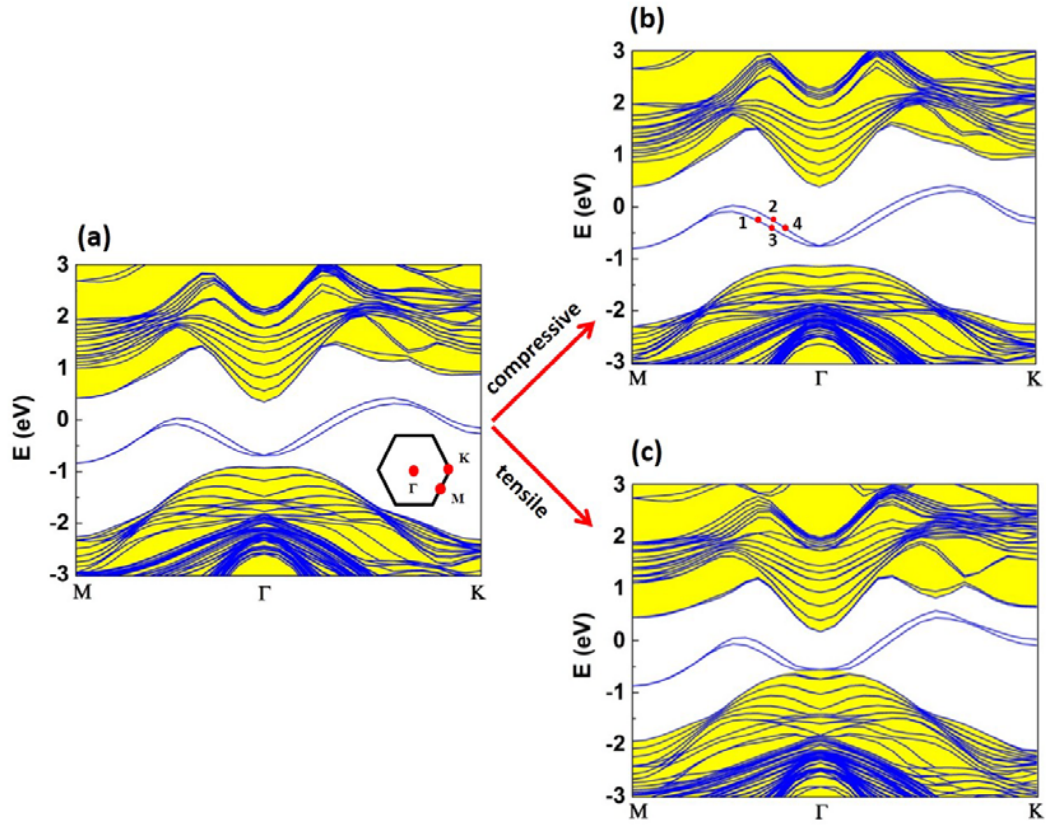

**Figure 2.** Band structures of Au/InSe(0001) under different substrate strains: (a) strain-free lattice constant $a$ = 4.05 Å. The inset is the schematic of the 2D hexagonal Brillouin Zone for Au/InSe(0001). (b) Compressive strain = -1.2% at $a$ = 4.00 Å. (c) Tensile strain = +2.4% at $a$ = 4.15 Å. In (b), 1&2 label two representative Rashba states with energy E = -0.28 eV and 3&4 with E = -0.40 eV.

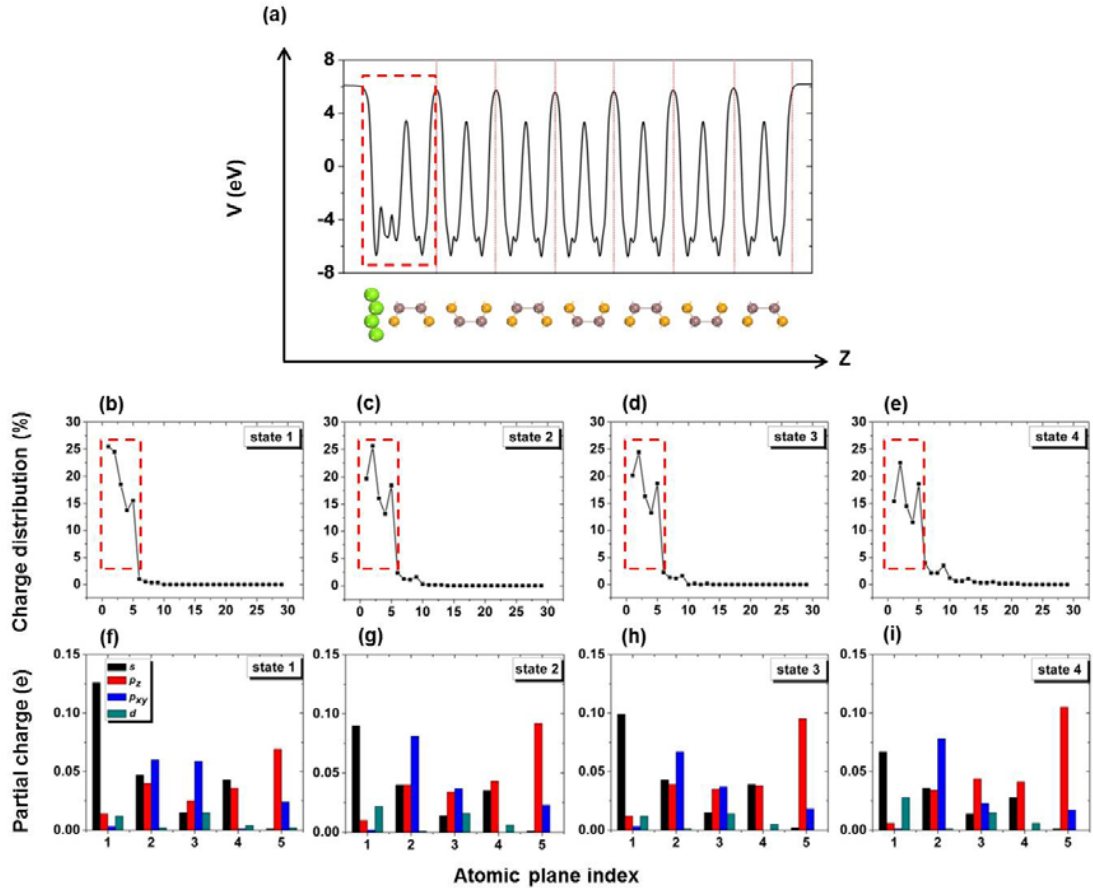

**Figure 3.** (a) Plane-averaged potential V along z-axis. (b-e) Charge distribution of the Rashba states 1-4, as labeled in Fig. 2(b) as as a function of atomic plane. (f-i) Orbital compositions in the first five atomic planes for the Rashba states 1-4. The red-dashed boxes highlight V only at the first five atomic planes. The red dotted lines in (a) represent the vdW spacing positions between InSe(0001) layers.

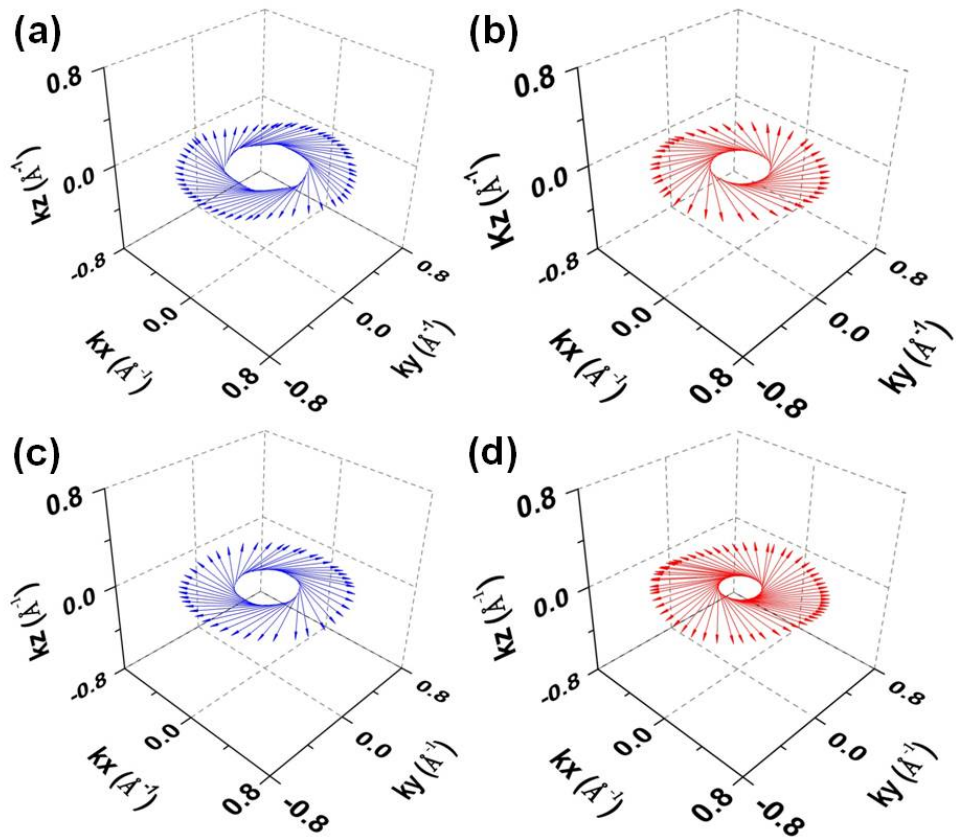

**Figure 4.** Spin polarization textures of the Rashba bands at 2D iso-energy arcs with E = -0.28 eV (a-b) and E = -0.40 eV (c-d), respectively. (a,c) are from the outer circle of corresponding iso-energy arc, and (b,d) are from the inner circle of corresponding iso-energy arc.

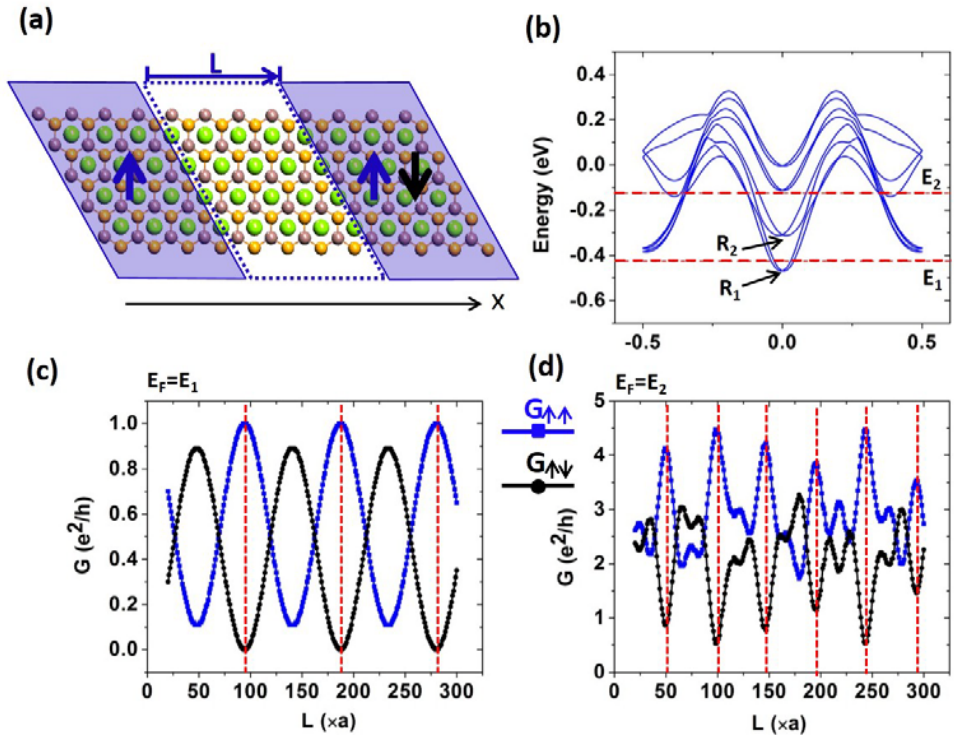

**Figure 5.** (a) Datta-Das transistor: the left and right boxes represent the source and drain leads without SOC, respectively. Both leads are ferromagnetic with their spins (shown in arrows) aligned either parallel or antiparallel. The middle box represents the conducting channel with SOC. (b) Band structure of the nanoribbon channel. (c, d) The spin-up $G_{\uparrow\uparrow}(E_F)$ and spin-down $G_{\uparrow\downarrow}(E_F)$ as a function of channel length (L) in units of lattice constant *a*. The vertical dashed lines label the conductance peak positions.

# Supporting Information

# Formation of Ideal Rashba States on Layered Semiconductor Surfaces Steered by Strain Engineering

*Wenmei Ming,* [†,‡] *Z. F. Wang,* [†,§] *Miao Zhou,* [†,∥] *Mina Yoon,* [*‡] *and Feng Liu* [*†,±]

[†]Department of Materials Science and Engineering, University of Utah, Salt Lake City, Utah 84112, United States

[‡]Center for Nanophase Materials Sciences, Oak Ridge National Laboratory, Oak Ridge, Tennessee 37831, United States

[§]Hefei National Laboratory for Physical Sciences at the Microscale, University of Science and Technology of China, Anhui 230026, China

[∥] Key Laboratory of Optoelectronic Technology and Systems of the Education Ministry of China, College of Optoelectronic Engineering, Chongqing University, Chongqing 400044, China

[±]Collaborative Innovation Center of Quantum Matter, Beijing 100084, China

*E-mail: myoon@ornl.gov (M.Y.)

*E-mail: fliu@eng.utah.edu (F.L.)

**Table of contents:**

**I: Effects of strained substrate InSe(0001) on the Rashba bands**

**II: Discussion of Rashba splitting in tight-binding picture**

**III: Wannier tight-binding band structures**

## I: Effects of strained substrate InSe(0001) on the Rashba bands

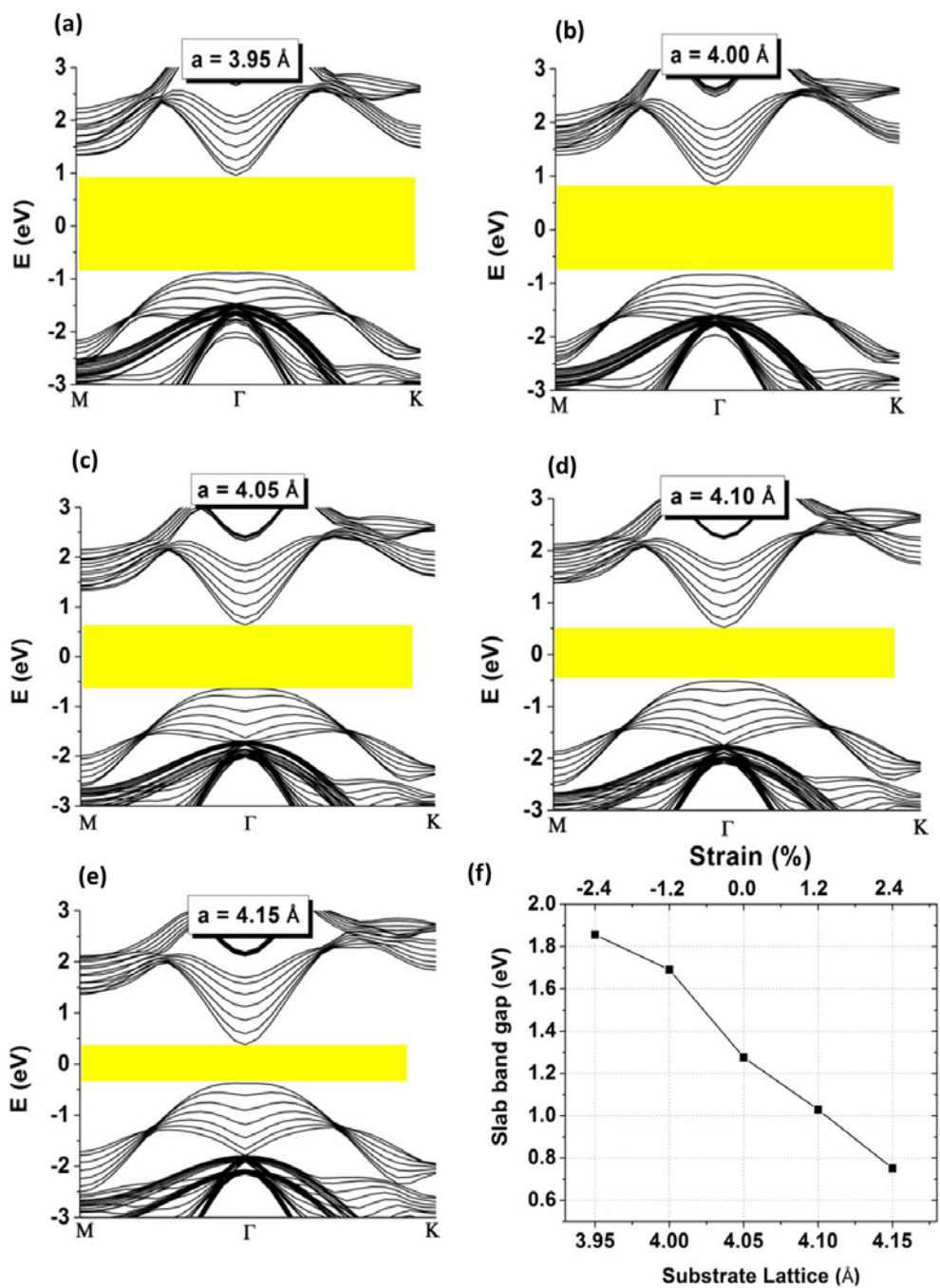

**Figure S1.** (a-e) HSE06 band structure of InSe(0001) slab at different lattice constants *a* from 3.95 Å to 4.15 Å. (f) Band gap versus substrate lattice constant *a*.

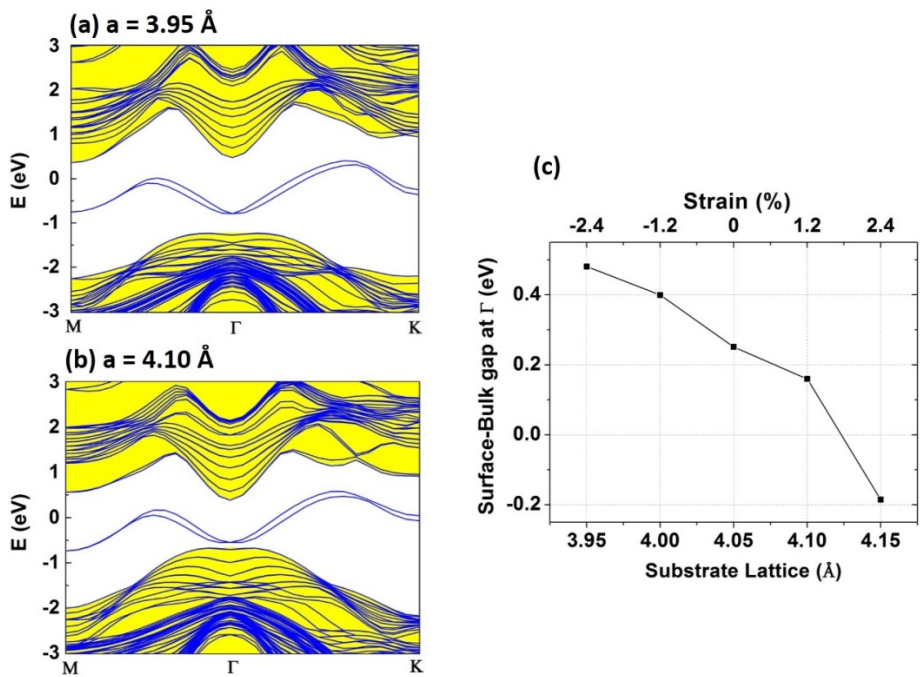

**Figure S2.** (a-b) HSE06 band structure for Au/InSe(0001) at substrate lattice constant $a$ = 3.95 Å and 4.10 Å. (c) Energy separation between Rashba bands and substrate valence band maximum (VBM) at Γ point as a function of substrate lattice constant

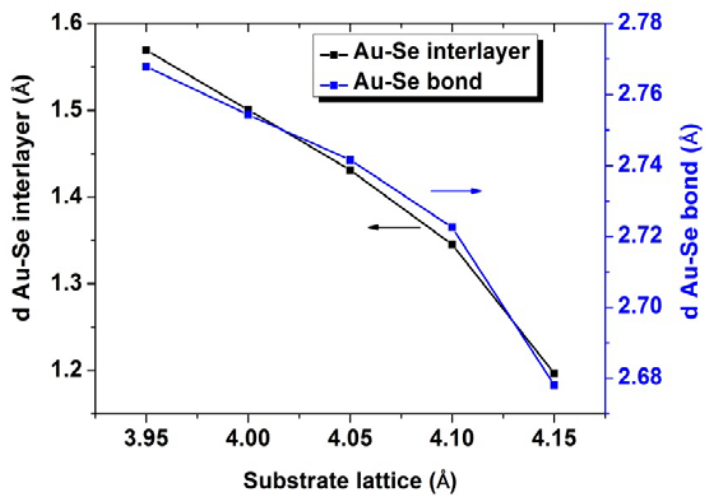

**Figure S3.** (Left axis) Vertical interlayer distance between Au and Se of InSe(0001) and (right axis) Au-Se bond length as a function of substrate lattice constant $a$.

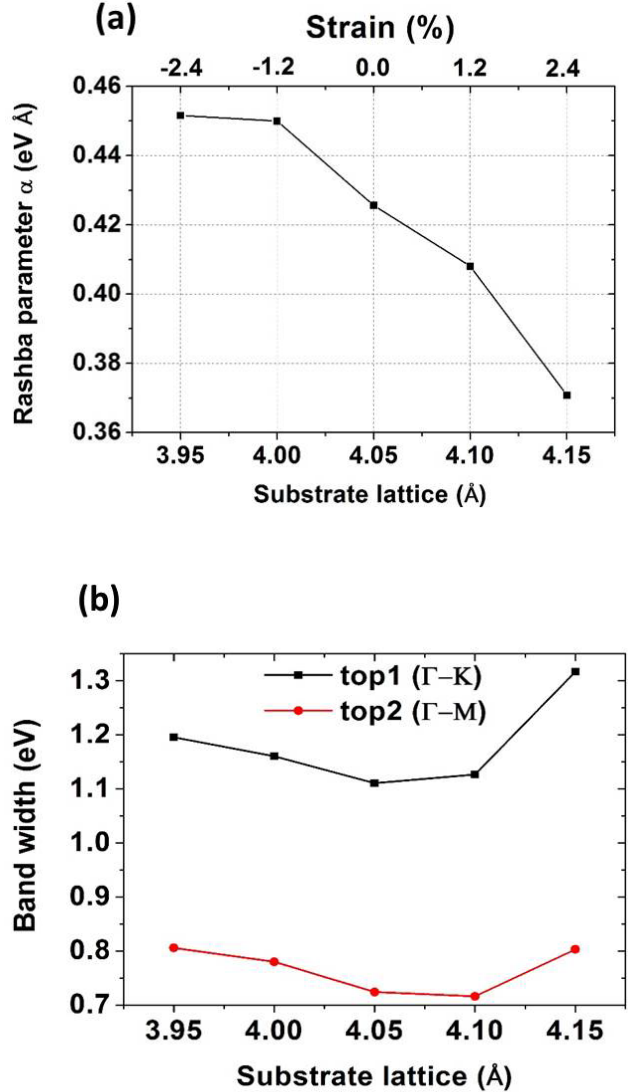

**Figure S4.** (a) Variation of Rashaba parameter α with substrate lattice constant (strain). (b) Variation of Rashba band width with substrate lattice constant, where top1 represents the bandwidth along Γ-K direction and top2 represents the bandwidth along Γ-M direction.

In (a) we see that Rashba parameter α decreases with increasing substrate lattice constant. When the substrate lattice constant increases, the effective in-plane Au nearest-neighbor *s-s* spin-flip hopping is expected to decrease, and hence results in decreasing Rashba parameter. In (b) we observe that before the overlap between the Rashba surface bands and the substrate bands happens, the bandwidths show slight decrease with increasing substrate lattice constant. When these two groups of bands get very close, the bandwidths show increase probably due to stronger interaction between Au overlayer and substrate in the perpendicular direction. Overall the change of band dispersion and bandwidth under strain is not significant.

## II: Discussion of Rashba splitting in tight-binding picture

In tight-binding picture, Rashba effect effectively represents intersite spin-flip hopping in presence of a perpendicular potential asymmetry and SOC. Generally such intersite spin-flip hopping can be induced between orbitals (l, m) by finite onsite coupling with $\Delta l = \pm 1$ and $\Delta m = 0$ and by intersite coupling with $\Delta l = 0$ and $\Delta m \neq 0$, under perpendicular potential asymmetry. The former process was seen for Rashba effect in gated graphene, where one has to include all the s, $p_x$, $p_y$, $p_z$ orbitals to induce Rashba spin splitting of graphene π band, even though graphene π band is of ~100% $p_z$ character, because otherwise $p_z$ alone would not have any SOC (see reasons in table I). The latter process was seen for Rashba effect in Au(111), where one has to include $p_x$, $p_y$ and $p_z$ orbitals (see table I). See also references (*Phys. Rev. B 82, 245412 (2010); Sur. Sci. 459, 49 (2000)*). To better see the two effective spin-flip hopping processes, we sketch them for gated graphene in Figure S5(a) and Au(111) in Figure S5(b) below:

**Table S1.** Matrix elements of atomic SOC Hamiltonian $H = L \cdot S$ in atomic basis of s, $p_x$, $p_y$ and $p_z$. It is obvious that spin flip can only occur between $p_x$ and $p_z$, or between $p_y$ and $p_z$.

| Orbitals | s | $p_x$ | $p_y$ | $p_z$ |
|---|---|---|---|---|
| s | 0 | 0 | 0 | 0 |
| $p_x$ | 0 | 0 | $-iS_z$ | $iS_y$ |
| $p_y$ | 0 | $iS_z$ | 0 | $-iS_x$ |
| $p_z$ | 0 | $-iS_y$ | $iS_x$ | 0 |

As shown in Fig. 3 f-i of our maintext, the four tested Rashba states have large amount of Au s-orbital character and s-orbital alone cannot give rise to spin-orbit coupling. However, they also contain small amount of Au $p_z$, $p_{xy}$ orbital characters, which is likely to be responsible for the appearance of SOC in Au/InSe(0001). Furthermore, notice that in tight-binding picture, Rashba effect effectively represents intersite spin-flip hopping in presence of a perpendicular potential asymmetry and SOC, so the SOC couplings between $p_z$, $p_x$ and between $p_z$ and $p_y$ have to be included, which are the only mechanisms of flipping spin (see the matrix elements of atomic SOC Hamiltonian $H = L \cdot S$ in the table I above).

We additionally provide the atomistic mechanism in Figure S5(c) below to illustrate the electron hopping processes involved to flip spin in Au/InSe(0001). Effectively they lead to electron hopping from spin-up s state to spin-down s state at a nearest-neighbor site, which is the very reason for strong Rashba splitting, The last process is onsite hopping from Au $p_z$ orbital to s orbital, which can only be possible under a finite perpendicular electric field and is quantified by nonzero matrix element $z_{sp}=\langle s|z|p_z\rangle$.

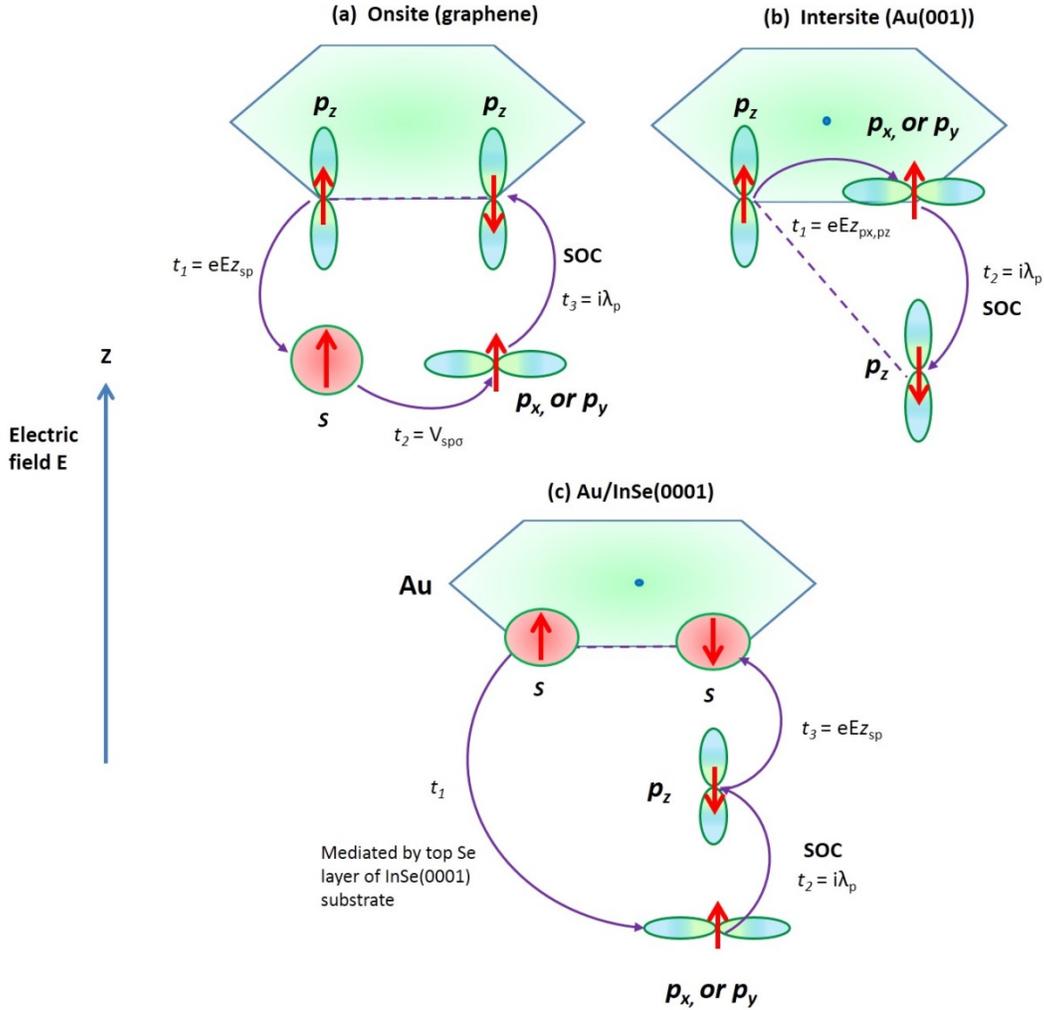

**Figure S5.** Two different mechanisms from electric field induced onsite contribution (a) [graphene] and intersite contribution (b) [Au(111)] for Rashba effect. (c) Microscopic electron hopping process to flip spin for generating Rashba effect in Au/InSe(0001). The solid purple lines represent the involved hopping processes. The amplitude of each process is denoted by $t_i$ (process index $i$ =1, 2 or 3), where $z_{sp}$=<s|z| $p_z$ > with s and $p_z$ at the same site. $\lambda_p$ is the spin-orbit coupling (SOC) strength of p orbitals. The arrows represent spins. For clarity, the different orbitals on the same site are placed along vertical direction z.

## III: Wannier tight-binding band structures

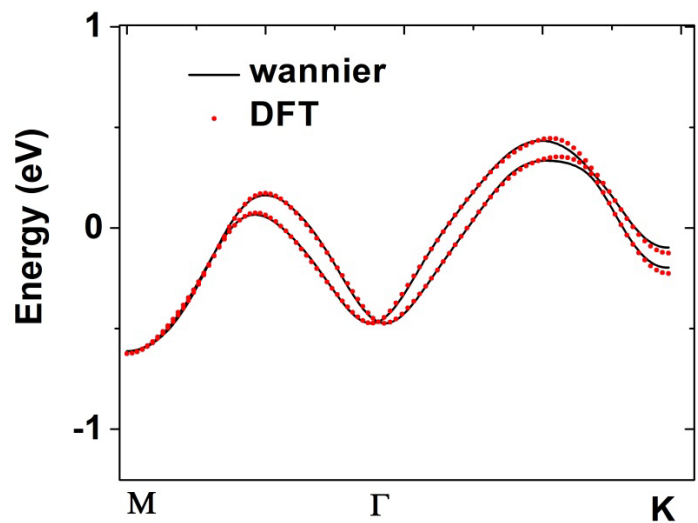

**Figure S6.** Comparison of Au/InSe(0001) band structures from DFT and Wannier fitting. It shows good match between them.

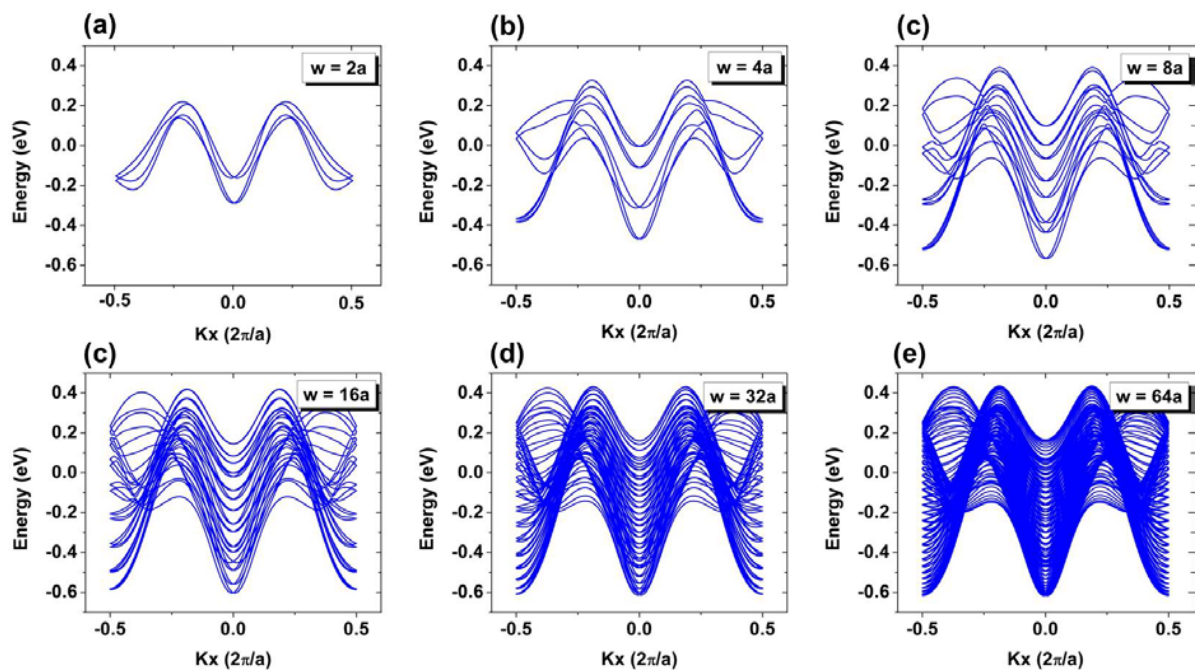

**Figure S7.** Wannier tight-binding band structures of Au/InSe(0001) nanoribbon with different width w.